\documentclass
[aps,preprint,showkeys,a4paper,tightenlines,superscriptaddress]{revtex4}%
\usepackage{eurosym}
\usepackage{amsfonts}
\usepackage{amsmath}
\usepackage{amssymb}
\usepackage{graphicx}
\usepackage{subfig}
\usepackage{caption}%
\providecommand{\U}[1]{\protect\rule{.1in}{.1in}}
\providecommand{\U}[1]{\protect\rule{.1in}{.1in}}
\providecommand{\U}[1]{\protect\rule{.1in}{.1in}}
\providecommand{\U}[1]{\protect\rule{.1in}{.1in}}
\providecommand{\U}[1]{\protect\rule{.1in}{.1in}}
\newtheorem{theorem}{Theorem}
\newtheorem{acknowledgement}[theorem]{Acknowledgement}

\begin{document}
\title{Tuning the periodic V-peeling behavior of elastic tapes applied to thin
compliant substrates}
\author{N. Menga}
\email{nicola.menga@poliba.it [Corresponding author]}
\affiliation{Imperial College London, Department of Mechanical Engineering, Exhibition
Road, London SW7 2AZ}
\affiliation{Department of Mechanics, Mathematics and Management, Politecnico of Bari, V.le
Japigia, 182, 70126, Bari, Italy}
\author{D. Dini}
\affiliation{Imperial College London, Department of Mechanical Engineering, Exhibition
Road, London SW7 2AZ}
\author{G. Carbone}
\affiliation{Department of Mechanics, Mathematics and Management, Politecnico of Bari, V.le
Japigia, 182, 70126, Bari, Italy}
\affiliation{CNR - Institute for Photonics and Nanotechnologies U.O.S. Bari, Physics
Department \textquotedblright M. Merlin\textquotedblright, via Amendola 173,
70126 Bari, Italy}

\begin{abstract}
In this paper, we investigate the periodic peeling behavior of opposing
symmetric peeling fronts involving an elastic tape peeled off from a
deformable substrate of finite thickness, backed onto a rigid foundation.

We treat the problem by means of an energetic formulation, and we found that,
depending on the values of the initial detached length $l$, substrate
thickness $h$, and peeling periodicity $\lambda$, the translational invariance
of the peeling process is lost and restored, as the elastic interaction
between the peeling fronts is limited by the substrate thickness. Indeed,
given $h$ and $\lambda$, a critical value of the detached length can be found,
which is able to prevent unstable peeling of the tape under a fixed applied
load, thus resulting in enhanced adhesion strength, with respect to the
classical Kendall's solution for peeling from a rigid substrate. On the other
hand, given the geometrical system configuration (i.e. the detached length
$l$) the load necessary to trigger the peeling can be minimized by
conveniently tuning the ratio $h/\lambda$. This feature might be of interest
for the development of innovative designs for future biomedical devices, such
as Transdermal Drug Delivery Systems or wound dressing, requiring low peel
adhesion for safe successive removals.

\end{abstract}

\copyright{ 2019. This manuscript version is made available under the
CC-BY-NC-ND 4.0 license http://creativecommons.org/licenses/by-nc-nd/4.0/}

\maketitle

\section{Introduction}

Nowadays, it is well established that peeling process of thin tapes and
membrane is one of the most important detachment mechanism, both in natural
and human-related applications. Moving from the pioneeristic studies by
Kaelble \cite{Kaelble1959,Kaelble1965} and Kendall \cite{Kendall1975}, who set
the thermodynamic framework to describe peeling processes, it is now clear
that the interplay between adhesive interfacial interactions and elastic
deformations of the contacting bodies plays a key role on the overall peeling
behavior \cite{VarembergPugnoGorb2010, ding2001, Shanahan2000, Wan2001,
Jin2009, Pesika2007, Pugno2011}. Aiming at understanding and, possibly,
mimicking their extraordinary adhesive performance, several experimental
observations \cite{Autumn2002, Autumn2006, Huber2005} on arachnids and
reptiles toes have, indeed, suggested that hierarchically arranged structures
of hairs or spatulae may play a key role in the attachment and detachment
process. Theoretical studies \cite{Carb2011,AffCarb2013,Persson2003}, and
experimental bio-inspired investigations
\cite{Geim2003,ArtzGorb2003,Glassmaker2007,Murphyetal2011,delCampo2007} have,
then, confirmed that very compliant hierarchical structures entail large
intimate contact area between the contacting bodies, thus resulting in high
van der Waals short-range adhesive interactions even in the case of rough
substrates \cite{Ozer2016,ORorke2016,creton,martina}.

However, although ensuring enhanced adhesive performances, these structures
also allow for easy detachment, thus allowing the animals to rapidly walk and
climb. To this regard, successive investigations
\cite{Tian2006,Autumn-etal2006,Gravish2008,PugnoPugno2012}, specifically
devoted to the attachment and detachment mechanics of gecko toes, have shown
that the detachment process is governed by peeling, and that a non-negligible
interplay exists between the toe adhesive performance and the peeling angle of
the multiple spatulae and fibrils, thus leading to both high adhesive strength
and low peel adhesion. Similarly, the effect of friction on adhesion has been
experimentally assessed, showing that interfacial shear stress nonlinearly
affect the pull-off load and may even slightly increase the adhesive
performance of thin fibrillar pads \cite{Varenberg2007}, as indeed predicted
in a recent theory \cite{Menga2018}.

Following these studies, several attempt have been made to mimic this combined
high adhesive-low peeling behavior in industrial applications. For instance,
several climbing robots \cite{Krahnetal2011,Murphyetal2011} equipped with
bio-inspired attachment pads
\cite{delCampo2007,Lee2008,Pugno2008,carbone2012sticky,Sekiguchi2015} have
been developed, all exploiting the superior adhesion provided by very
compliant mushroom- and hairy-shaped end structures. Peeling was then employed
to easily control the detachment process of such systems.

However, most of the existing studies, only focus on the case of elastic thin
structures, hierarchically arranged, in adhesive contact with significantly
stiffer substrates, thus usually considered as rigid. Although this assumption
holds true for a large class of practical applications, peculiarly demanding
problems exist in which the effect of the substrate elasticity on the peeling
behavior cannot be neglected. It is the case, for instance, of thin protective
peelable packaging for food and general soft components, in which the peeled
tape is stiffer than the substrate.

Of course, dealing with a single elastic tape peeled away from an elastic
substrate would not affect Kendall's results \cite{Kendall1975}, due to the
translational invariance of the elastic field in the substrate, which
therefore does not contribute to the energy balance of the advancing peeling.
However, for instance, if we assume a viscoelastic behavior for, at least, one
among the tape and the substrate the peeling behavior may turn out
significantly altered, as due to viscous dissipation in the materials the
process is now velocity dependent. It is the case, for instance, of the
peeling of a viscoelastic tape from a rigid substrate \cite{Peng2014}, in
which increasing peeling force with the peeling rate is reported. Similarly,
the more complex case of an elastic tape peeled away from a viscoelastic
adhesive has been firstly treated in Ref. \cite{AffCarb2016}, where enhanced
adhesive toughness has been reported, and then in Ref. \cite{Perrin2019},
where, in the framework of linear viscoelasticity, the authors define specific
scaling laws for the process.

Also the number of peeling fronts involved in the peeling process, as well as
the specific spatial arrangement of them, may strongly affect the peeling
performances. Moving from natural observation of biological systems
\cite{Wolff2015,Heepe2017} (\textit{e.g.} web anchors), a specific adhesion
mechanism has been identified which relies on multiple pattern of opposing
double-sided peeling fronts (V-peeling). Both theoretical \cite{Pugno2011} and
experimental \cite{Wolff2017} studies have indeed shown that dealing with
V-peeling from rigid substrates, both geometrical and hierarchical
optimizations can be performed in order to increase the maximum sustainable
load. Further investigations on the same V-peeling configuration have
addressed the case of uniform cohesive interactions at the interface
\cite{Gialamas2014}, also taking into account for the tape large deformations
and frictional sliding effects \cite{Begley2013} which, indeed, significantly
increase the pull-off load.

Definitely less has been done in the case of V-peeling from deformable
substrates. Of course, in this case, also the elastic energy stored in the
substrate has to be taken into account in the overall energy balance governing
the peeling evolution, since, as the peeling advances and the fronts moves
apart from each other, the elastic field in the substrate changes. A first
attempt to investigate such peculiar scenario has been made in Ref.
\cite{Bosia2014} by relying on computationally demanding Finite Element (FE)
simulations. However, a more comprehensive investigation on this topic has
been recently provided in Ref. \cite{Menga2018Vpeeling} where, by relying on
reliable Boundary Element (BE) calculations, the case of periodic V-peelings
from an elastic half-space is treated, eventually finding that a
non-negligible interactions between the multiple peeling fronts occurs through
the elastic fields in the substrate. This entails that the results strongly
depend on the distance between the peeling fronts, thus making the process no
longer translationally invariant.

To this regard, it is important to observe that there exist a large class of
applications in which the elasticity of the substrate is crucial in
determining the overall adhesive performance of the system.\emph{ }Among the
others, this is the case of biomedical applications such as pressure sores
dressing and Transdermal Drag Delivery Systems (TDDS). Several attempts have
been made to analyze in details the peel adhesion performance of these
Pressure Sensitive Adhesives (PSA)
\cite{Chivers2001,Karwoski2004,Renvoise2009,Bait2013}, however the usual
half-space assumption falls short in describing the role played by the
substrate deformability in the adhesive behavior for such applications.
Indeed, for instance, TDDS usually stuck on sufficiently thin biological
tissues confined by the underlaying stiffer bones. Since high holding power
and low peel adhesion are essential requirements for the medical functionality
of such systems \cite{Minghetti2004,Wokovich2006,Cilurzo2012}, the possibility
to explore innovative designs relying on periodically arranged V-peeling
fronts in order to reduce the peeling resistance can be important for future
developments. To this regard, a significant effort has been made, in both
elastic and viscoelastic contact mechanics, to clarify that, in case of bodies
of finite thickness, an additional lengthscale (\textit{i.e.} the body size)
competes in defining the system behavior
\cite{carb-mang-2008,Menga2016,Menga2016visco}, leading to non-negligible
effects in terms of contact stiffness and area.

Here, we present a study on the physical behavior of the multiple periodically
distributed double V-shaped opposing peeling of an elastic thin tape from an
elastic substrate of finite thickness, backed onto a rigid foundation, a
configuration that mimics the above scenarios. We perform our study in the
framework of linear elasticity, considering quasi-static reversible
thermodynamic transformations, in the same path already drawn by Kendall
\cite{Kendall1975}. The displacement field in the elastic substrate is solved
by means of a Green's function approach, by exploiting the formalism recently
given in Ref. \cite{Menga2019coupling}. The results here presented provide an
improved understanding of the effect of the geometrical parameters on the
peeling behavior of periodic arrangements, which can potentially be used to
minimize/maximize the peeling load for different applications.

\section{Formulation}

\begin{figure}[ptbh]
\centering\includegraphics[width=1\textwidth]{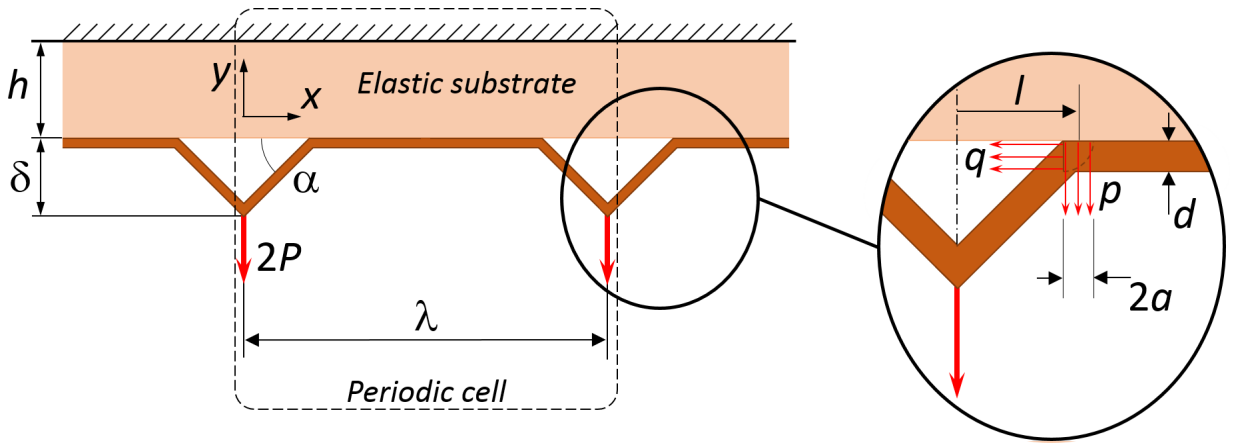}\caption{The system
configuration under periodic V-peeling of an elastic tape of thickness $d$
from an elastic substrate of thickness $h$.}%
\label{fig1}%
\end{figure}

Figure \ref{fig1} shows a sketch of the system under investigation: a thin
elastic tape of thickness $d$ is periodically peeled away from an elastic
layer of thickness $h$, which is backed onto a rigid foundation. The
detachment occurs via multiple V-shaped opposite peeling fronts under periodic
normal loads $2P$, whose spatial periodicity is $\lambda$. Since in each
periodic cell, both the loads and system geometry are symmetric, we will
conveniently focus our study on half of the periodic cell. We can easily
define the overall load on each peeling front as $F=P/\left(  \sin
\alpha\right)  $, with $\alpha$ being the peeling angle between the tape and
the substrate. Further, by defining $\varepsilon=F/\left(  Ebd\right)  $ the
tape elongation and $l$ the detached length, simple geometrical arguments
\cite{Menga2018Vpeeling} show that $l\left(  1+\varepsilon\right)  \cos
\alpha=l$, which finally gives the following implicit expression of the
peeling angle $\alpha$%
\begin{equation}
P-Ebd\left(  \tan\alpha-\sin\alpha\right)  =0 \label{1}%
\end{equation}
where $E$ is the Young modulus of elasticity of the tape,\ and $b$ is the
transverse width.

In this work, the description of the peeling behavior relies on the energetic
formulation firstly proposed in Refs
\cite{Kaelble1959,Kaelble1965,Kendall1975}; thus we focus on linear
elasticity, neglecting any large deformation effect. Of course, dealing with
very soft materials, large deformations effects might slightly change the
quantitative evolution of the process; however, as shown in previous works on
similar topics \cite{JKR1971,Lin2006}, the main physical picture would not be
affected, and qualitative inferences can still be drawn from linear theory.
Similarly, detailed studies have shown that dynamic effects may significantly
alter the peeling behavior, also giving rise to instabilities during the
peeling evolution
\cite{Dembo1988,Ciccotti2004,Rumi2006,Dalbe2014a,Dalbe2014b,Qian2017}.
However, since our study aims at investigating the fundamental mechanisms
linked to geometric effects (\textit{e.g.} thickness), here we focus on
quasi-static conditions, thus neglecting any dynamic effect (\textit{i.e.} we
assume that the peeling fronts propagation occurs far slower that the speed of sound).

As already observed in Ref. \cite{Menga2018Vpeeling}, since here the tape is
peeled away from an elastic substrate via a periodic V-peeling process, the
geometrical configuration changes as the detachment fronts move apart. This
implies that the elastic field within the substrate is no longer stationary,
thus the translational invariance typical of Kendall's peeling is lost. Under
these conditions, the elastic energy term associated with the deformation of
the substrate have to be taken into account in defining the total potential
energy of the system $U_{tot}$. Therefore we have
\begin{equation}
U_{tot}=U_{el,t}+U_{el,s}+U_{P}+U_{ad} \label{2}%
\end{equation}
where $U_{el,t}$ and $U_{el,s}$ are the elastic energy stored in the tape and
in the substrate, respectively. $U_{P}$ is the potential energy associated
with the normal load $P$, and $U_{ad}$ is the adhesion energy. We assume the
elastic tape sufficiently thin to neglect any bending contribution to its
deformation (see Ref \cite{Villey2017}). The resulting deformation is of pure
stretch, thus $U_{el,t}=F^{2}l/\left(  2Ebd\right)  $. Similarly, the surface
adhesion energy is $U_{ad}=lb\Delta\gamma$, where $\Delta\gamma$ is the work
of adhesion, also known as the Dupr\`{e} energy of adhesion.

For what concerns the energetic term associated to the deformation of the
elastic substrate, building on the physical arguments already given in Refs.
\cite{AffCarb2016,Menga2018Vpeeling,Kaelble1960}, we assume that the load $F$
acting on each branch of the V-shaped tape is balanced by a uniform
distribution of normal $p=F\sin\alpha/\left(  2ab\right)  $ and tangential
$q=\pm F\cos\alpha/\left(  2ab\right)  $ tractions localized at the tip of the
detachment fronts over a region of size $2a\approx d$ (see fig. \ref{fig1}).
Indeed, very thin tapes show vanishing bending stiffness ($\propto d^{3}$),
thus leading to highly concentrated normal interfacial stresses (see Refs.
\cite{Kaelble1960,Lin2007}); furthermore, dimensional arguments show that
tangential tractions must interest a region of size $\left(  E/E_{s}\right)
^{1/2}\sqrt{\lambda d}$, with $E_{s}$ corresponding to the substrate's elastic
modulus. Although the assumption of $2a\approx d$ may be restrictive in some
case, we expect this not to affect our assessment of the physical scenario
under consideration, provided that $2a+l\ll\lambda/2$.

Since the substrate is thin, we expect the total interfacial displacement
$\mathbf{u}^{t}\left(  x\right)  =\left(  u_{x}^{t},\text{ }u_{y}^{t}\right)
$ to be function of both the layer thickness $h$ and loads periodicity
$\lambda$ \cite{Menga2019coupling,MengaRLRB2017,Menga2018rough}. In order to
calculate $\mathbf{u}^{t}\left(  x\right)  $, we conveniently rely on the
superposition of the periodic displacement fields $\mathbf{u}^{r,l}\left(
x\right)  $ independently related to the $p$ and $q$ stresses of each of the
two adjacent peeling fronts ($r,l$ indicate for the \textit{right} and
\textit{left} detachment fronts). From Fig. \ref{fig1}, it follows that
$\mathbf{u}^{t}\left(  x\right)  =\mathbf{u}^{r}\left(  x-l\right)
+\mathbf{u}^{l}\left(  -x-l\right)  $, where $\mathbf{u}^{r,l}$ can be found
by relying on a Green's function approach based on the fundamental periodic
solutions given in Ref. \cite{Menga2019coupling} for the case of an elastic
layer of finite thickness backed onto a rigid foundation (see Appendix A).
Finally, the elastic energy stored in the substrate can be found as
\begin{equation}
U_{el,s}=\frac{1}{2}b\int_{l-a}^{l+a}\mathbf{\sigma}\cdot\mathbf{u}%
^{t}\text{d}x \label{3}%
\end{equation}
being $\mathbf{\sigma}=\left(  q,\text{ }p\right)  $.

In the same framework, the potential energy associated to the applied vertical
load $P$ is given by%

\begin{equation}
U_{P}=-P\left[  u_{y}^{t}\left(  l\right)  +\delta\right]  \label{4}%
\end{equation}
where $u_{y}^{t}\left(  l\right)  $ is the substrate normal displacement of
the point of application of $P$, and $\delta=l\tan\alpha$.

Following Ref. \cite{maugis}, for isothermal reversible transformations, the
peeling process equilibrium requires that%
\begin{equation}
\frac{\partial U_{tot}}{\partial A}=\frac{1}{b}\frac{\partial U_{tot}%
}{\partial l}=0 \label{5}%
\end{equation}
where $A$ is the detached tape area.

Since, from Eq. (\ref{2}), $U_{tot}$ is sensitive to both the vertical load
$P$ and the detached length $l$, Eq. (\ref{5}) allows to determine whether,
for a given value of the controlled parameter $P$, a specific value of $l$
exists able to ensure equilibrium.

\section{Results}

Several works have shown that, dealing with elastic substrates of finite
thickness in periodic contacts, a peculiar role is played by the ratio
$h/\lambda$, as the thickness $h$ represents a limit to the elastic
interaction length \cite{MengaRLRB2017,Menga2018rough}. In the case of rough
contacts, for instance, this implies that asperities whose distance is larger
that $h$ do not interact with each other. For this reason, we do expect the
ratio $h/\lambda$ to significantly affect the V-peeling behavior, thus we
focus out study on different values of $h$ and $\lambda$.

Furthermore, several experimental studies on human-related soft tissues for
biomedical applications have reported an almost incompressible behavior of the
tissues \cite{Liang2009,Li2011}, thus here we focus on a substrate with
Poisson's ratio $\nu_{s}=0.5$. Regarding the tape, we assume $d=10^{-4}$ m,
$E=730$ MPa, and $b=0.015$ m, as in several of the available commercial
adhesive tapes. We define the elasticity ratio $\chi=E/E_{s}$ between the tape
and substrate elastic moduli, and the dimensionless peeling load $\tilde
{P}=P/P_{0}$, with $P_{0}$ being the Kendall's peeling load in the case of
rigid substrates \cite{Kendall1975}.

\begin{figure}[ptbh]
\centering\subfloat[\label{fig2a}]{\includegraphics[width=0.5\textwidth]{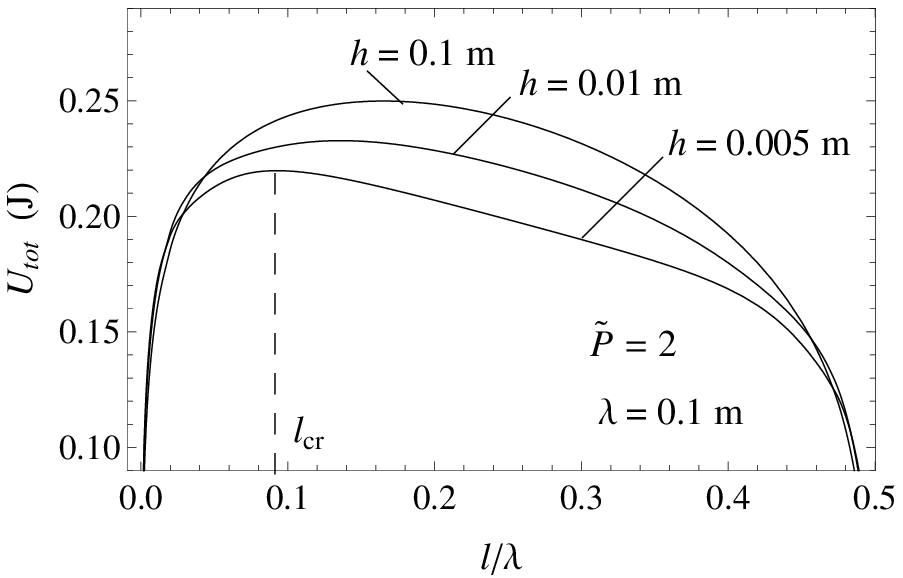}}
\hfill\subfloat[\label{fig2b}]{\includegraphics[width=0.5\textwidth]
{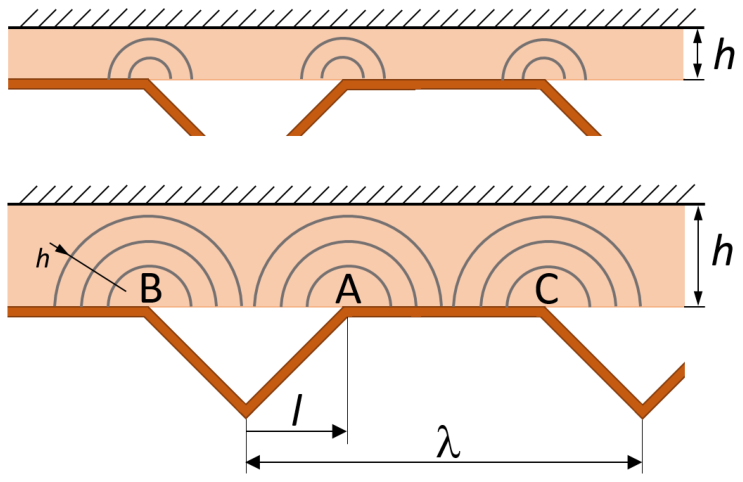}}\caption{(a) The total energy $U_{tot}$ as a function of the
dimensionless detached length $l/\lambda$ for different values of the subtrate
thickness, at fixed peeling load $P/P_{0}=2$. Calculations refer to
$\lambda=0.1$ m, and $\Delta\gamma=74$ J/m$^{2}$. (b) A physical scheme of the
interaction between adjacent peeling fronts. }%
\label{fig2}%
\end{figure}

In Fig. \ref{fig2a}, the total energy $U_{tot}$ is plotted against the
dimensionless detached length $l/\lambda$, at given dimensionless peeling load
$\tilde{P}=2$. A critical detached length $l_{cr}$ exists where the
equilibrium condition of Eq. (\ref{5}) is satisfied. However, since
$\partial^{2}U_{tot}/\partial l^{2}<0$, $l_{cr}$ represents an unstable
equilibrium condition \cite{maugis}. This means that, for $l>l_{cr}$ the
system is no longer able to sustain the load, and unstable peeling propagation
occurs up to complete detachment; on the contrary, for $l<l_{cr}$, peeling is
prevented and, in the ideal case of perfectly reversible process, a
self-healing reattachment would be observed. Of course, in real applications,
several sources of dissipation can be identified in the proximity of the
peeling front (\textit{i.e.} non-conservative adhesive bonds braking, high
elastic strain rate, dynamic effects) able to prevent the self-healing
behavior, which indeed reduces to simple non-advancing peeling for $l<l_{cr}$.
Furthermore, in Fig. \ref{fig2a} we can also appreciate the effect of the
elastic layer thickness $h$. Specifically, by focusing on the thinner layer,
we observe that significant interaction between the peeling fronts through the
deformable substrate occurs only for sufficiently small, and large, values of
$l$. Indeed, dimensional arguments indicate that, due to the finiteness of the
elastic substrate thickness, the elastic fields associated to each of the
peeling fronts can only affect a circular region of the elastic substrate of
radius $r\approx h$. The mechanism of such an interaction is clearly shown in
top sketch in Fig. \ref{fig2b}: for $l\lessapprox h$, peeling fronts A and B
interact, whereas for $l\gtrapprox\lambda/2-h$, cracks A and C interact
instead. In the intermediate range of value of $l$, all the peeling fronts are
independent of each other, thus resulting in local translational invariance of
the substrate deformation, and in turn in a linear trend of the total energy
(i.e constant \textit{energy release rate}), as predicted by the Kendall
theory \cite{Kendall1975}. Notably, under this condition, peeling propagation
would occur for $\tilde{P}>1$. Increasing the layer thickness leads to a
different scenario (see the bottom sketch in Fig. \ref{fig2b}), as for
$h\gtrapprox\lambda$ all the peeling fronts are able to mutually interact
regardless of the value of $l$.

\begin{figure}[ptbh]
\centering\subfloat[\label{fig3a}]{\includegraphics[width=0.5\textwidth]{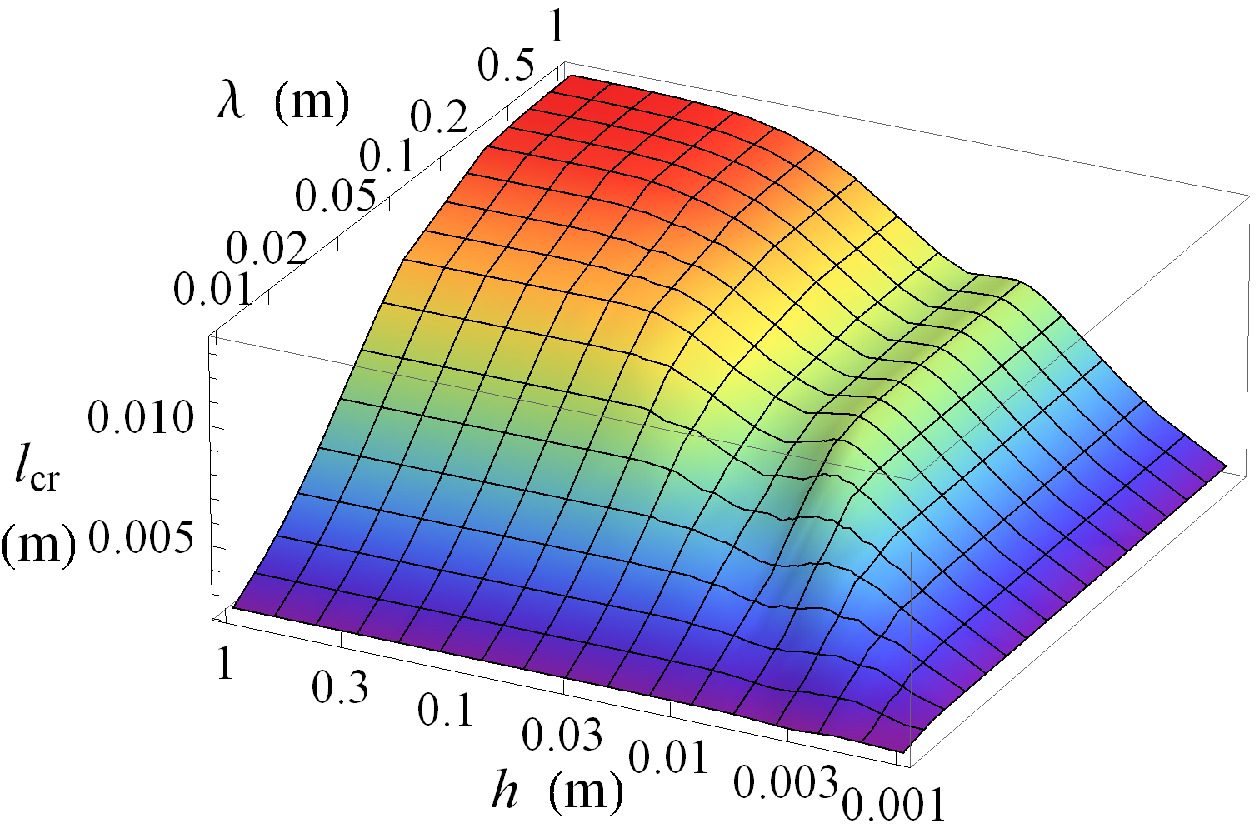}}\hfill
\subfloat[\label{fig3b}]{\includegraphics[width=0.45\textwidth]{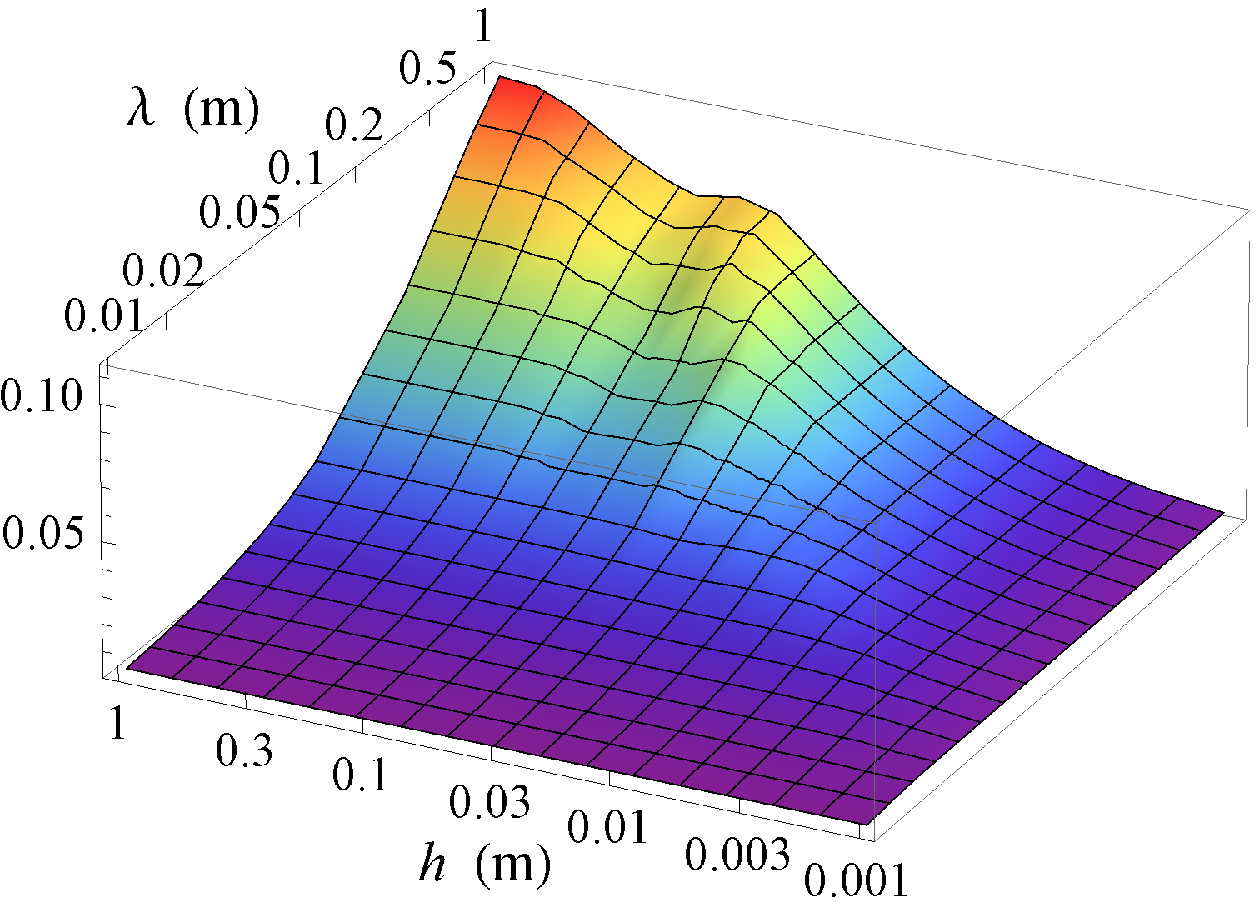}}\caption{Surface
representations of the effect of the peeling spatial periodicity $\lambda$ and
substrate thickness $h$ on the critical detached length $l_{cr}$, at fixed
load $P/P_{0}=1.1$. Two different elasticity ratios have been considered: (a)
$\chi=100$, and (b) $\chi=1000$.}%
\label{fig3}%
\end{figure}

The effect of the system configuration in terms of both the spatial
periodicity $\lambda$ of the peeling loads and the substrate thickness $h$ on
the critical detached length $l_{cr}$ is shown in Figs. \ref{fig3}, for two
different values of the elasticity ratio $\chi$. Regardless of the values of
$\lambda$ and $h$, we observe that the softer the substrate, the tougher the
peeling, as very compliant substrates enhance the peeling fronts interaction,
thus leading to larger critical detached length before peeling occurs
(\textit{i.e.} larger $l_{cr}$ at given load). Further, increasing $\lambda$
and $h$ leads to the asymptotic non-periodic half-space peeling behavior (see
Ref. \cite{Menga2018Vpeeling}). Interestingly, a non-monotonic trend of
$l_{cr}$ with respect to $h$ is also reported.

\begin{figure}[ptbh]
\centering\subfloat[\label{fig4a}]{\includegraphics[width=0.5\textwidth]{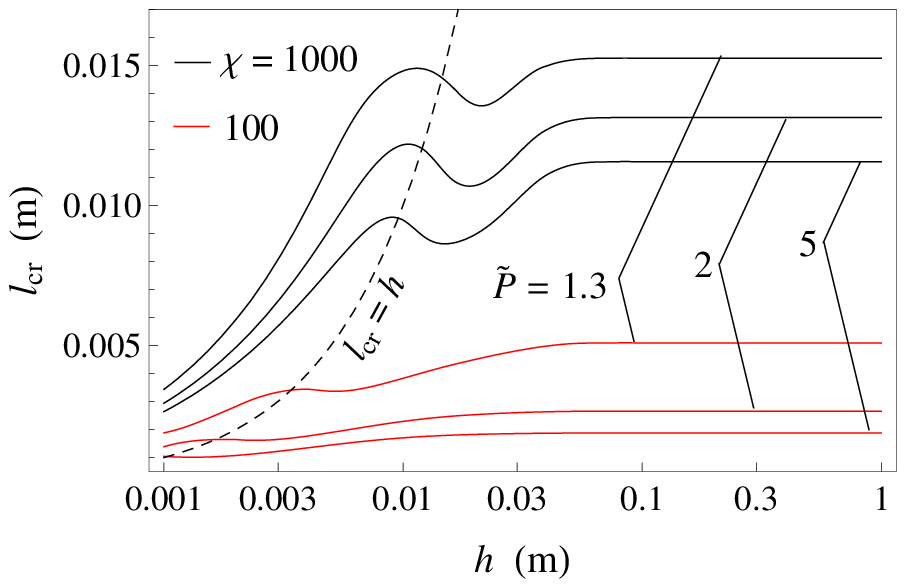}}
\hfill\subfloat[\label{fig4b}]{\includegraphics[width=0.45\textwidth]
{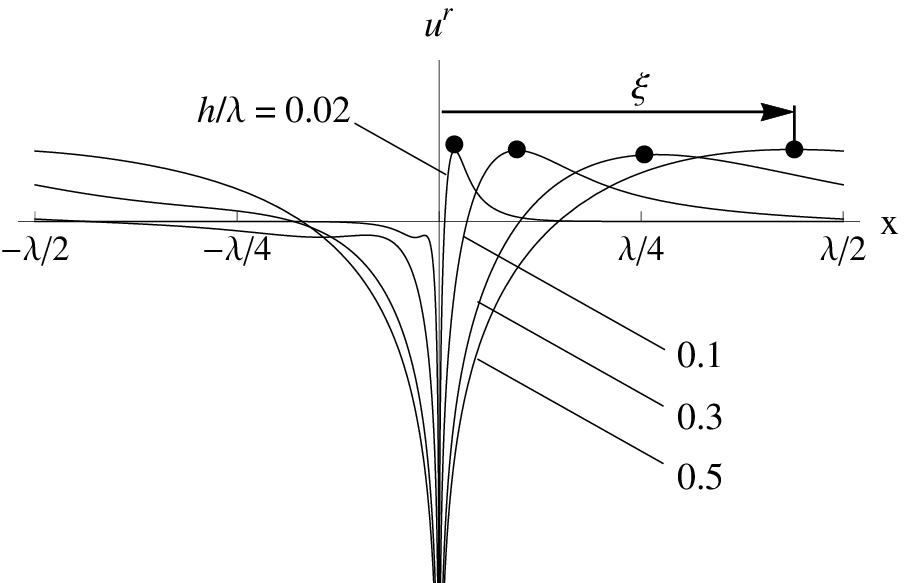}}\caption{(a) The non monotonic trend of the critical detached
length $l_{cr}$ as a function of the substrate thickness $h$, for different
values of the peeling load $P$ and elasticity ratio $\chi=E/E_{s}$. (b) the
general trend of normal displacements of the elastic substrate induced by
generic $p=q>0$ (see Fig. \ref{fig1})\ tractions for different substrate
thickness $h$.}%
\label{fig4}%
\end{figure}

Such a peculiar behavior is even more evident in Fig. \ref{fig4a} where the
critical detached length $l_{cr}$ is shown as a function of the substrate
thickness $h$ for different peeling loads and elasticity ratios $\chi$. In
order to qualitatively explain this interesting feature of interacting
multiple peeling process involving deformable substrate of finite thickness,
in Fig. \ref{fig4b} we plot the general trend of normal displacement field of
the substrate for different values of $h/\lambda$ due to a distribution of
normal and tangential tractions such as those related to the single crack A of
Fig. \ref{fig2b}. We observe that reducing $h$ the size $\xi$ of the
significantly deformed region of the layer surface reduces too, as $\xi\approx
h$ \cite{Menga2019coupling}. Indeed, given the detached length of the
V-peeling process (\textit{i.e.} the distance between the two peeling fronts A
and B in Fig. \ref{fig2b}), due to elastic interactions, the peeling fronts
will affect each other differently depending on the substrate thickness $h$,
thus modifying both the amount of elastic energy stored in the substrate, and
the overall potential energy associated to the normal peeling load $P$ (see
Eqs. (\ref{3}-\ref{4})). Consequently, in Fig. \ref{fig4a} we note that,
regardless of $\tilde{P}$ and $\chi$, this non-monotonic effect occurs
approximately at $l_{cr}\approx h$, where the displacement fields induced by
the opposing peeling fronts present the highest degree of interaction with
each other. Finally, reducing $\chi$ leads to less pronounced effects of $h$,
as, given the normal load, stiffer substrates entail smaller displacements and
in turn smaller amount of elastic energy stored (\textit{i.e.} it selectively
affects only the terms $U_{P}$ and $U_{el,s}$ in Eq. (\ref{2})).

\begin{figure}[ptbh]
\centering\subfloat[\label{fig5a}]{\includegraphics[width=0.5\textwidth]{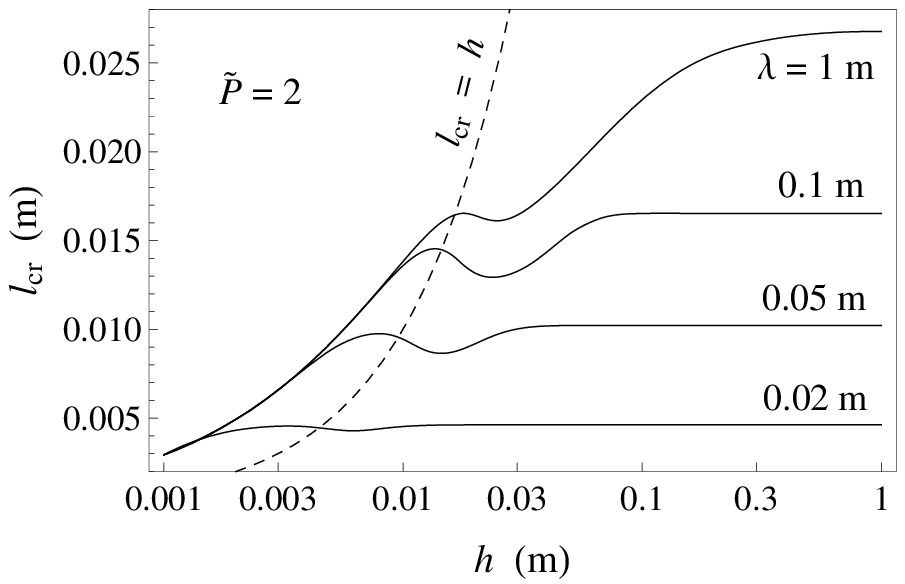}}
\hfill\subfloat[\label{fig5b}]{\includegraphics[width=0.5\textwidth]
{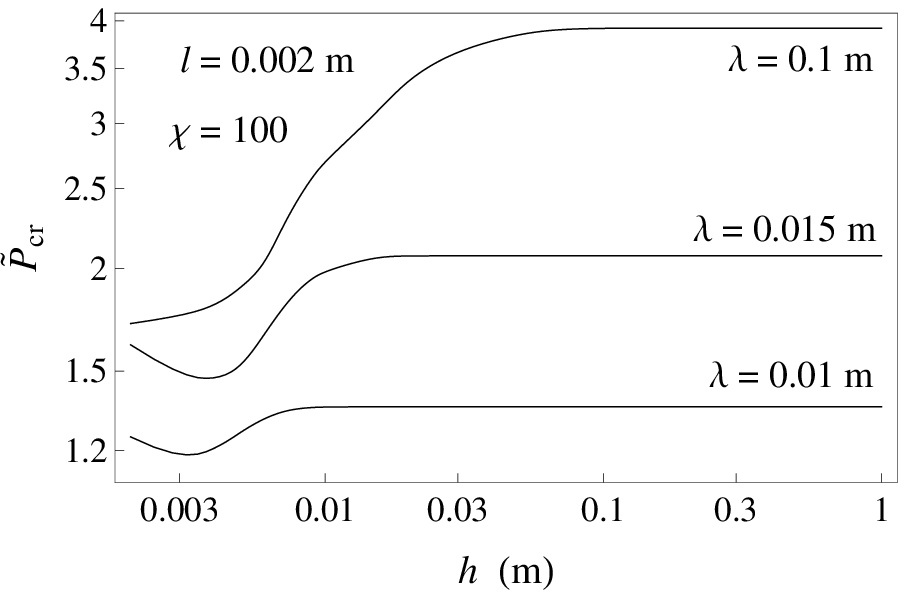}}\caption{(a) The critical detached length $l_{cr}$ under fixed
load, and (b) the dimensionaless critical peeling load $\tilde{P}_{cr}$
correspondig to a given detached length $l$, as functions of the substrate
thickness $h$, for different values of the system spatial periodicity
$\lambda$. Calculations refers to $\chi=100$.}%
\label{fig5}%
\end{figure}

In Fig. \ref{fig5a} we show the effect of the spatial periodicity $\lambda$ of
the periodic V-peelings on the peeling toughness, as indeed $l_{cr}$ is
plotted against $h$ for different value of $\lambda$. We observe that, given
the peeling load, increasing $\lambda$ entails larger detached length
tolerable before unstable peeling propagation. In other words, the interaction
between the fronts A and C in Fig. \ref{fig2b} reduces the overall peeling
toughness, whereas the interaction between the V-peeling fronts (A and B) is
beneficial in terms of higher tolerable detached length. Of course, per each
value of the layer thickness $h$, a further periodicity increase beyond
$\lambda\gtrapprox h$ does not change the physical scenario, as the asymptotic
aperiodic behavior is approached.

In Fig. \ref{fig5b}, instead, we show the effect of the $\lambda$ and $h$ on
the system in terms of the critical peeling load $\tilde{P}_{cr}$ above which
unstable peeling propagation occurs, given a value of $l$. In this case, we
observe that the peeling load can be minimized by tuning the substrate
thickness $h$ and system periodicity $\lambda$. Specifically, the reduction of
both $\lambda$ and $h$ leads to a reduction of $\tilde{P}_{cr}$. However,
depending on the value of $\lambda$, non trivial results are found, as a
global minimum of $\tilde{P}_{cr}$ as a function of $h$ is reported. These
results show how the V-peeling configuration can be exploited for innovative
optimization strategies to tailor the peel adhesion of several systems. It is
the case, for instance, of both wound dressing and TDDS in biomedical medical
applications, whose critical peeling load might be conveniently reduced by
relying on an innovative V-peeling removal mechanism, which adjusts the value
of $\lambda$ given a certain skin thickness $h$, thus reducing the possibility
of injuries in successive removals.

\section{Conclusions}

In this paper, we investigate the peeling behavior of a thin elastic tape
peeled off from an elastic substrate of finite thickness. The peeling process
occurs by means of symmetrically V-peeling fronts, periodically distributed
over the substrate. Due to the specific system configuration, the
translational invariance of the peeling process is lost and restored,
depending on the specific value of the detached length, substrate thickness,
and spatial periodicity.

As a consequence, the overall results show a certain dependency on these
parameters, in terms of both peel-off load and toughness. In general, we found
that the sustainable peeling load depends on the detached length; however, the
ratio between the substrate thickness and spatial periodicity affects this
trend, giving rise to local minima. Such a peculiar feature offers the
opportunity to conveniently tune both the substrate thickness and spatial
peeling periodicity in order to minimize the peeling load. We hope this
research provides the fundamental understanding needed to develop new adhesive
solutions, possibly opening new design opportunities to enhance the peeling
behavior of, for instance, wound dressing and TDDS in biomedical applications,
as well as on general purpose packaging films.

\begin{acknowledgement}
This project has received funding from the European Union's Horizon 2020
research and innovation programme under the Marie Sk\l odowska-Curie grant
agreement No 845756 (N.M. Individual Fellowship). D.D. acknowledges the
support received from the Engineering and Physical Science Research Council
(EPSRC) through his Established Career Fellowship EP/N025954/1.
\end{acknowledgement}

\appendix{}

\section{Elastic surface displacements of thin layers under periodic uniform
normal and tangential tractions}

\begin{figure}[ptbh]
\centering\includegraphics[width=0.7\textwidth]{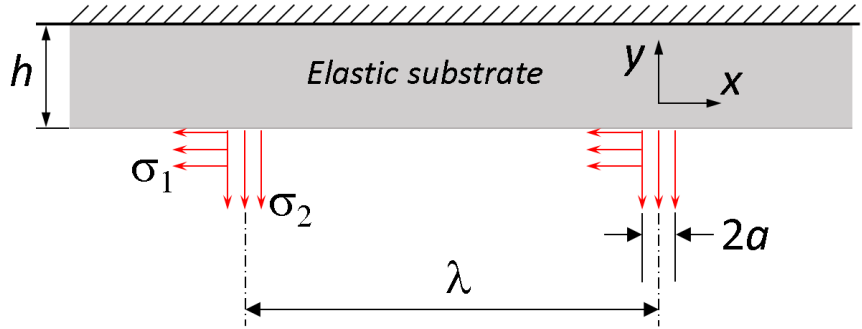}\caption{The
elastic problem at hand: an elastic layer of thickness $h$ is loaded with a
periodic distribution of uniform normal and tangential tractions over a strip
of size $2a$. The spatial periodicity of the loads is $\lambda$.}%
\label{fig1app}%
\end{figure}

We focus on the problem shown in Fig \ref{fig1app} of an elastic substrate of
thickness $h$ subjected to a distribution of uniform normal $\sigma_{2}$ and
tangential $\sigma_{1}$ tractions acting over a strip of size $2a$.

Further, we assume the tractions $\sigma_{1}$ and $\sigma_{2}$ to be
periodically distributed, with periodicity $\lambda$. Following Ref.
\cite{Menga2019coupling}, by means of the the convolution product between the
Green's tensor $\mathbf{G}\left(  x\right)  $ and the interfacial stress
vector $\mathbf{\sigma=}\left(  \sigma_{1},\sigma_{2}\right)  $ (see Fig.
\ref{fig1app}), one can find the total surface displacement vector
$\mathbf{v}=\left(  v_{1},v_{2}\right)  $ as%
\begin{equation}
\mathbf{v}\left(  x\right)  -\mathbf{v}_{m}=\int_{\Omega}ds\mathbf{G}\left(
x-s\right)  \mathbf{\sigma}\left(  s\right)  ;\text{ \ \ \ \ \ \ }x\in
\lbrack0,\lambda] \label{v-vm}%
\end{equation}
where $\Omega$ is the domain of integration (\textit{i.e.} $\Omega=[-a,a]$),
$\mathbf{v}_{m}$ is the mean plane displacement vector and the components of
the the Green's tensor are given by \cite{Menga2019coupling}%
\begin{align}
G_{11}\left(  x\right)   &  =\frac{2\left(  1-\nu_{s}^{2}\right)  }{\pi E_{s}%
}\left[  \log\left\vert 2\sin\left(  \frac{kx}{2}\right)  \right\vert
+\sum_{m=1}^{\infty}A_{m}\left(  kh\right)  \frac{\cos\left(  mkx\right)  }%
{m}\right]  ,\label{G11}\\
G_{12}\left(  x\right)   &  =-G_{21}\left(  x\right)  =\frac{1+\nu_{s}}{\pi
E_{s}}\left[  \frac{1-2\nu_{s}}{2}\left[  \text{sgn}\left(  x\right)
\pi-kx\right]  -\sum_{m=1}^{\infty}B_{m}\left(  kh\right)  \frac{\sin\left(
mkx\right)  }{m}\right]  ,\label{G12}\\
G_{22}\left(  x\right)   &  =\frac{2\left(  1-\nu_{s}^{2}\right)  }{\pi E_{s}%
}\left[  \log\left\vert 2\sin\left(  \frac{kx}{2}\right)  \right\vert
+\sum_{m=1}^{\infty}C_{m}\left(  kh\right)  \frac{\cos\left(  mkx\right)  }%
{m}\right]  . \label{G22}%
\end{align}
where $k=2\pi/\lambda$, and the quantities $A_{m}$, $B_{m}$, and $C_{m}$ are
given by%
\begin{align}
A_{m}\left(  kh\right)   &  =1-\frac{2khm+\left(  3-4\nu_{s}\right)
\sinh\left(  2khm\right)  }{5+2\left(  khm\right)  ^{2}-4\nu\left(  3-2\nu
_{s}\right)  +\left(  3-4\nu_{s}\right)  \cosh\left(  2khm\right)  }%
,\label{A}\\
B_{m}\left(  kh\right)   &  =\frac{4\left(  1-\nu_{s}\right)  \left[
2+\left(  khm\right)  ^{2}-6\nu_{s}+4\nu_{s}^{2}\right]  }{5+2\left(
khm\right)  ^{2}-4\nu_{s}\left(  3-2\nu_{s}\right)  +\left(  3-4\nu
_{s}\right)  \cosh\left(  2khm\right)  }.\label{B}\\
C_{m}\left(  kh\right)   &  =1+\frac{2khm-\left(  3-4\nu_{s}\right)
\sinh\left(  2khm\right)  }{5+2\left(  khm\right)  ^{2}-4\nu\left(  3-2\nu
_{s}\right)  +\left(  3-4\nu_{s}\right)  \cosh\left(  2khm\right)  },
\label{C}%
\end{align}

By recalling that we refer to a uniform distribution of tractions, after
substituting Eqs. (\ref{G11}-\ref{G22}), Eq. (\ref{v-vm}) can be conveniently
rewritten in terms of the interfacial displacements $\mathbf{u}=\mathbf{v}%
\left(  x\right)  -\mathbf{v}_{m}=\left(  u_{1},u_{2}\right)  $%
\begin{align}
u_{1}\left(  x\right)   &  =q\left[  Z_{11}\left(  x+a\right)  -Z_{11}\left(
x-a\right)  \right]  +p\left[  Z_{12}\left(  x+a\right)  -Z_{12}\left(
x-a\right)  \right] \label{u1}\\
u_{2}\left(  x\right)   &  =-q\left[  Z_{12}\left(  x+a\right)  -Z_{12}\left(
x-a\right)  \right]  +p\left[  Z_{22}\left(  x+a\right)  -Z_{22}\left(
x-a\right)  \right]  \label{u2}%
\end{align}
\ where the functions $F_{1}\left(  x\right)  $ and $F_{2}\left(  x\right)  $
are so defined%
\begin{align}
Z_{11}\left(  x\right)   &  =\frac{2\left(  1-\nu_{h}^{2}\right)  }{\pi
E_{h}k}\left(  -Cl_{2}\left(  kx\right)  +\sum_{m=1}^{\infty}A_{m}\left(
kh\right)  \frac{\sin\left(  mkx\right)  }{m^{2}}\right) \\
Z_{12}\left(  x\right)   &  =\frac{1+\nu_{s}}{\pi E_{s}k}\left(  \frac
{1-2\nu_{s}}{2}\Phi\left(  kx\right)  +\sum_{m=1}^{\infty}B_{m}\left(
kh\right)  \frac{\cos\left(  mkx\right)  }{m^{2}}\right) \\
Z_{22}\left(  x\right)   &  =\frac{2\left(  1-\nu_{h}^{2}\right)  }{\pi
E_{h}k}\left(  -Cl_{2}\left(  kx\right)  +\sum_{m=1}^{\infty}C_{m}\left(
kh\right)  \frac{\sin\left(  mkx\right)  }{m^{2}}\right)
\end{align}
being
\begin{equation}
Cl_{2}\left(  x\right)  =-\int_{0}^{x}\log\left\vert 2\sin\left(  t/2\right)
\right\vert dt
\end{equation}
the Clausen function of order $2$ (Ref. \cite{Abramowitz}), and%
\begin{equation}
\Phi\left(  x\right)  =\int_{0}^{x}\left[  \text{sgn}\left(  t\right)
\pi-t\right]  dt
\end{equation}

\end{document}